# Computational design of high performance hybrid perovskite on silicon tandem solar cells

A. Rolland[1], L. Pedesseau[1,a], A. Beck[1], M. Kepenekian[2], C. Katan[2], Y. Huang[1], S. Wang[1], C. Cornet[1], O. Durand[1], and J. Even[1,a]

[1] *UMR FOTON, CNRS, INSA Rennes, Rennes, F35708, France*

[2] *Institut des Sciences Chimiques de Rennes, UMR 6226, CNRS, Université de Rennes 1, 35042 Rennes, France*

[a] *Corresponding authors. laurent.pedesseau@insa-rennes.fr (L. Pedesseau), jacky.even@insa-rennes.fr (J. Even)*

**Abstract:**

In this study, the optoelectronic properties of a monolithically integrated series-connected tandem solar cell are simulated. Following the large success of hybrid organic-inorganic perovskites, which have recently demonstrated large efficiencies with low production costs, we examine the possibility of using the same perovskites as absorbers in a tandem solar cell. The cell consists in a methylammonium mixed bromide-iodide lead perovskite, $CH_3NH_3PbI_{3(1-x)}Br_{3x}$ ($0 \leq x \leq 1$), top sub-cell and a single-crystalline silicon bottom sub-cell. A Si-based tunnel junction connects the two sub-cells. Numerical simulations are based on a one-dimensional numerical drift-diffusion model. It is shown that a top cell absorbing material with 20% of bromide and a thickness in the 300-400 nm range affords current matching with the silicon bottom cell. Good interconnection between single cells is ensured by standard n and p doping of the silicon at $5.10^{19} cm^{-3}$ in the tunnel junction. A maximum efficiency of 27% is predicted for the tandem cell, exceeding the efficiencies of stand-alone silicon (17.3%) and perovskite cells (17.9%) taken for our simulations, and more importantly, that of the record crystalline Si cells.

## 1. INTRODUCTION

In order to ensure a large part of the worldwide energy production, photovoltaics has to increase its watt produced over price ratio and match, if not exceed, the one of nuclear and fossil energies. In recent years, two strategies have emerged from this objective. The first path consists in increasing significantly the solar-cell efficiencies. In that sense, one solution is to develop group III-V multi-junction solar cells (MJSC), which combine absorbers with different bandgap energies allowing to harvest a wider range of the solar spectrum while lowering intrinsic losses such as thermalisation losses [1,2], thus surpassing the well-known Schockley-Queisser limit of an ideal single-junction solar cell [3]. Such multigap systems have achieved conversion efficiencies exceeding 40% [4]. However, these MJSC are currently developed on expensive GaAs and Ge substrates and require high-quality materials, which incurs a substantial associated cost. Targeting terrestrial applications, less expensive alternatives have been proposed with two junction terminal cells containing a silicon (Si) bottom sub-cell, and developed on a low-cost Si substrate [5–8]. For a Si bottom sub-cell with a bandgap of 1.12 eV, it has been shown that the maximum conversion efficiency are achieved with a top sub-cell bandgap lying between 1.6 and 1.8 eV [9,10].

Following a different strategy, hybrid organic-inorganic perovskite (HOP) materials, such as lead methylammonium tri-iodide $CH_3NH_3PbI_3$, in so-called 'perovskite cells' have recently attracted huge attention [6,11–16]. These solar cells have demonstrated remarkably high efficiencies (around 20%) and high open-circuit voltages ($V_{oc}$) [17], but also offer great potential for low costs of production. Taking advantage of both these strategies, this class of material is a suitable candidate as a top sub-cell absorber for tandem cells built on CIGS or Si [18,19]. In this context, design of the tandem cell may further benefit from technological progress already achieved in other sectors, such as the use of a tunnel junctions (TJ), or Esaki diode [20], to connect electrically both sub cells that has already been implemented in III-V multigap devices [21–24], For instance, an all-Si made interband TJ offers a low resistance behavior due to the high doping values reachable in Si [18,19]. Early theoretical papers based on density functional theory, have shown that a strong spin-orbit coupling effect splits the conduction band of both $CH_3NH_3PbI_3$ and $CH_3NH_3PbBr_3$ compounds [25]. Simplified effective mass approaches and solid state physics concepts can be adapted to hybrid perovskite compounds and heterostructures [26,27].

Although it is a key point in the optimization of device architecture, only few simulations have been reported on perovskite-based solar cells [15,28]. As a major feature, the behavior of photo-carriers in HOP materials can be dealt with in the same way than in inorganic materials [15]. This similarity enables the use of existing drift-diffusion based device simulator widely used in the context of conventional semiconductors.

In this work, we conduct numerical simulations of two-terminal tandem cells based on HOP materials as top sub-cell absorber monolithically integrated on a bottom Si sub-cell. The sub-cells are electrically connected through a low-resistance

TJ. We chose $CH_3NH_3PbI_{3(1-x)}Br_{3x}$ ($0 \leq x \leq 1$, where x defines the composition) as top sub-cell absorber material. In fact, it has recently been shown that introducing a mixture of bromide and iodide in the halide composition of lead methylammonium perovskites allows for continuous fine-tuning the band gap [29]. We will show that this is important to ensure proper current matching between top and bottom cells. Besides, halide concentration, the thickness of the absorber influences the overall performances of the total cell. This effect will be carefully analyzed. Our computational study provides clear guidance to design an optimal HOP based tandem cell. The best efficiency is obtained for a bromide ratio x=0.2 and a thickness of the HOP layer between 300 and 400 nm. The tandem cell exhibits a conversion efficiency of 27%, significantly higher than the efficiencies of the stand-alone Top and Bottom sub-cell chosen for our simulations, which amount to 17.9% and 17.3%, respectively.

## 2. DEVICE STRUCTURE AND SIMULATION METHOD

Simulations are performed using the Silvaco Atlas device simulator [30], which allows to numerically solve Poisson's equation coupled with the continuity equations for both electrons and holes under steady state conditions. It is possible to account for quantum effects, which are mandatory to simulate a TJ. Different levels of mesh are required for tandem cell devices in order to consider layer thicknesses that vary from tens of nanometers to hundreds of micrometers in realistic devices. Figure 1 shows a scheme of a perovskite tandem solar cell monolithically integrated on a *n*-doped silicon substrate. The structure consists of 850 nm thick hole transport material (HTM) layer, a $CH_3NH_3PbI_{3(1-x)}Br_{3x}$ absorbing layer of various thickness, and a 300 nm thick $TiO_2$ electron transport layer. The 2x20 nm thick $Si(n^{++})/Si(p^{++})$ TJ insures the electrical connection between the two subcells. A 280 μm thick *n*-type Si substrate ($N_D = 10^{16}$ cm$^{-3}$), which includes deep trap levels, with an energy level located at 0.5 eV above the valence band edge and a density corresponding to a diffusion length of 600 μm, Finally, a 100 nm thick $n^{++}$ Si layer ($N_D = 10^{19}$ cm$^{-3}$) is used for the back contact.

It is not obvious to define relevant values of the physical parameters needed to accurately simulate the total cell, particularly for the HOP materials since their growth procedures are not yet standardized. Table I summarizes the key physical parameters taken from the literature and used in our simulations for the HTM, HOP, $TiO_2$ and Si materials [15,28,31]. Following the work by Minemoto and Murata [15], HOP electron ($\mu_n$) and hole ($\mu_p$) mobilities are set to 2 cm$^2$V$^{-1}$s$^{-1}$. HOP materials are assumed to be *n*-type with a low doping level of $10^{13}$ cm$^{-3}$. Deep traps are also considered with an energy level located 0.5 eV below the conduction band edge and a density $N_T = 10^{15}$ cm$^{-3}$ [32,33], which corresponds to a carrier diffusion length of about 700 nm in good agreement with available experimental data [15,32].

The complex refractive index k for HTM, $TiO_2$ and $CH_3NH_3PbI_{3(1-x)}Br_{3x}$ layers was calculated from the absorption coefficient given by $\alpha = A\sqrt{(E - E_g)}$ where A is a prefactor, $E_g$ the material bandgap energy, and $k = \frac{1.24\alpha}{4\pi E}$ (with $\alpha$ in $cm^{-1}$ and $E = h\nu$ in eV). A is kept constant for all materials of interest. In the work by Minemoto and Murata [15,34], A amounts to $10^5$ $cm^{-1}eV^{-1/2}$ for all materials. Here, we just chose 2.5 $10^5$ $cm^{-1}eV^{-1/2}$ for the $CH_3NH_3PbI_{3(1-x)}Br_{3x}$ mixed halide, which leads to a better agreement with the experimental absorption coefficient reported in Ref. [34].

In addition to the thickness of the absorber, one essential parameter that can be optimized in a tandem cell is its bandgap. Here, the choice for the $CH_3NH_3PbI_{3(1-x)}Br_{3x}$ mixed halide allows covering a wide range of bandgaps from 1.55 eV (x=0) to 2.3 eV (x=1). Bandgap energies and electron affinities of mixed compositions were computed from linear interpolations of the values reported for the pure tri-iodide and tri-bromide compounds (Table I). In the following, the tandem cell efficiencies are computed for various bromide concentrations and various absorbing layer thicknesses.

## 3. EFFECT OF ABSORBER LAYER BANDGAP ENERGY AND THICKNESS

### 3.1. The current matching issue in the HOP/Si tandem cell

In this section, we first show that a good current matching between the Top and Bottom sub-cells is required to obtain a maximum efficiency of the tandem cell. It can be achieved thanks to bandgap tuning related to the halide composition of the HOP layer. Figure 2 provides the simulated current-voltage characteristics (J-V) of the top HOP sub-cell, the bottom silicon sub-cell and the HOP-based tandem cell under AM1.5 illumination, considering a mixed halide composition of x = 0.2. The doping level of the silicon TJ is fixed to $10^{20}$ $cm^{-3}$. These results are in good agreement with the literature [6,11–15,35]. For the silicon bottom cell, a high short-circuit current $J_{SC} = 52.2$ $mAcm^{-2}$ and an open-circuit voltage $V_{OC} = 0.57$ V are obtained and lead to a cell efficiency of 17.3%. The top HOP cell gives $J_{SC} = 25.0$ $mA.cm^{-2}$, $V_{OC} = 1.18$ V and a cell efficiency of 17.9%. For the tandem cell, the open-circuit voltage is $V_{OC} = 1.75$ V, which corresponds to the sum of the open-circuit voltages of each sub-cell. On the other hand, the short-circuit current is identical to that of the top cell, and corresponds to the maximum of the total current that can be reached in the tandem cell. It shows that in this device, the current limitation is mainly due to the HOP top sub-cell which can lead to a non-ideal current matching between top and bottom sub-cells.

It is known that the bandgap energy of the top sub-cell absorber material and the thickness of the absorbing layer are key parameters to obtain a good current matching between the two sub-cells [24]. Figure 3 shows the band diagram of the tandem cell at room temperature and at thermal equilibrium for various halide compositions of the HOP top sub-cell. Due to the low residual doping level of the HOP material, the absorbing layer acts almost as an intrinsic layer and the top sub-cell can be considered as a PIN junction. Moreover, whatever the halide composition, when an electron-hole pair is created in the perovskite layer, the electrons easily flow to the $TiO_2$ layer without encountering any potential barrier. The behavior of the holes is similar from the perovskite layer to HTM layer.

*3.2. Optimization of the HOP single cell*

In order to validate our simulations, we investigate the influence of the absorbing layer bandgap and its thickness on the HOP cell efficiency, and compare our results to available experimental data. Figure 4 shows the mapping of $J_{SC}$, $V_{OC}$ and efficiency (η) as a function of (*i*) the bromide concentration in the absorbing layer, and (*ii*) the thickness of the HOP layer in the solar cell. For a given thickness, the short circuit current $J_{SC}$ decreases when the bandgap energy increases because the amount of photo-generated carriers decreases. Concurrently, the open circuit voltage $V_{OC}$ increases. These contradictory trends lead to an optimum of the top sub-cell efficiency for an absorber bandgap energy of about 1.4 eV [36]. From our simulations, we find a maximum efficiency around 18.5% obtained for $CH_3NH_3PbI_3$ with a layer thickness between 300 and 500 nm. With a similar thickness we find an efficiency of 11.3% for the $CH_3NH_3PbBr_3$ compounds. These values compare well with the corresponding experimentally measured efficiencies of 17% ($CH_3NH_3PbI_3$) [37] and 9% ($CH_3NH_3PbBr_3$) [38,39] and show that our simulation is well suitable for further investigations of HOP tandem cells.

*3.3. Silicon tunnel junction*

The silicon tunnel junction used to interconnect both top and bottom sub-cell plays an important role and behaves as a short circuit when the tandem-cell operates [21,22,40]. Therefore, in order to optimize the whole device, the knowledge of the minimum doping level of the $n^{++}$ and $p^{++}$ region of this junction is mandatory for efficient band to band tunneling effect. Figure 5 shows the variation of the tunnel junction peak current and the differential negative resistance, which is representative of the peak-valley current ratio, as a function of the doping level of each region. The doping level is assumed to be the same on both side of the junction. A variation of more than six orders of magnitude is observed for an increase of the doping level from $3.10^{19}$ cm$^{-3}$ to $10^{20}$ cm$^{-3}$. For the latter value, the reverse tunnel junction current is greater than $10^5$ mA.cm$^{-2}$ under a reverse bias of 0.2 V.

Using the optimum values of HOP bandgap energy and layer thickness mentioned above, we simulate the properties of tandem cell for different values of the tunnel junction doping levels. The minimum doping level which leads to an efficient tunneling effect and a good interconnection between top and bottom sub-cells is around $5.10^{19}$ cm$^{-3}$. This value corresponds to the degeneracy limit of the material: below, the silicon material is no longer degenerated on the conduction band side, leading to a weak overlap of conduction and valence bands and poor band to band tunneling.

*3.4. Optimization of the HOP/silicon tandem-cell*

In this section we show that an optimum of the efficiency can be obtained for the HOP tandem cell by varying both the bandgap and thickness of the HOP absorbing layer. Indeed, starting from the optimized doping level of the TJ, we can optimize the complete tandem solar cell. Figure 6 shows the mapping of $J_{SC}$, $V_{OC}$ and η in a HOP/Si tandem cell as a function of (*i*) the bromide concentration in the absorbing layer, and (*ii*) its thickness. For a fixed bromide concentration x, *i.e.* a given HOP bandgap energy, $J_{sc}$ increases as a function of the thickness up to 1 µm. However, a small decrease in $J_{sc}$ has to be noted for thicker absorber due to a decrease of the electric field in the absorber layer and to carrier recombination effects. The trend of $V_{oc}$ variation as a function of the absorber thickness is a monotonous decrease whatever the value of the absorber bandgap energy. This is again due to the decrease of the electric field in the absorbing layer. This behavior is related to the balance between generation current in the absorber and the junction forward current. The cell efficiency presents an optimum for each value of the absorber bandgap energy. For the parameters used in the present simulation (Table I), an optimum as high as 27% is obtained for a bromide concentration of x=0.2 in the $CH_3NH_3PbI_{3(1-x)}Br_{3x}$ top sub-cell, which, corresponds to a bandgap energy of 1.7 eV, and a thickness between 300 and 400 nm of the HOP absorber layer. Noteworthy, this efficiency surpasses the record of the best silicon solar cell [35]. Table II summarizes the values of $J_{SC}$ and $V_{oc}$, fill factor (FF) and cell efficiency η computed for different halide compositions of the top sub-cell absorbing layer whose thickness is either 300 or 400 nm. The maximum value obtained for $J_{sc}$ is about 26.2 mA.cm$^{-2}$ for a HOP bandgap energy of 1.6 eV. The $V_{oc}$ increases from 1.6 V to 2.3 V when the absorber bandgap energy increases from 1.55 to 2.3 eV. Moreover, we find a maximum plateau of the efficiency for an absorber thickness from 300 nm to 400 nm.

The External Quantum Efficiency (EQE) can be calculated for the top and bottom cells following the procedure presented in Ref. [41]. Figure 7 shows the optimized EQE spectra of (*i*) the HOP top cell with bromide ratio x=0.2 and a thickness of 300 nm, and (*ii*) the Si bottom cell. At the optimum efficiency, complementary absorptions occur for the entire 300 to 1100 nm wavelength range, with balanced Jsc = 24.4 mA.cm$^{-2}$ and a $V_{oc}$ of 1.2 V for the HOP top cell and 0.55 V for the Si bottom cell. Figure 7 shows that the top cell is absorbing almost every wavelength under 0.72 µm, in reasonable agreement with experimental results [19].

In the present study, the limit between the two materials is very well defined due to the strong absorption of the HOP material. For bromide ratio higher than 0.2, the limit is shifted to lower wavelength, down to 0.55 µm for the pure $CH_3NH_3PbBr_3$. For high bromide ratio, the resulting tandem cell is less efficient, the bottom Si cell being less efficient than the HOP cell in this range of wavelength. Thus, the alloy perovskite absorber with a bromide ratio of 0.2 and a thickness in the 300-400 nm range insures the current matching condition of top and bottom cells.

## 4. CONCLUSION

In summary, based on the drift-diffusion model we have conducted a detailed investigation of hybrid organic-inorganic perovskite on silicon tandem solar-cells with a tunnel junction, including trap recombination effects. It is first shown that $5.10^{19}$ cm$^{-3}$ $p$ and $n$ doping levels of Si are high enough to guarantee appropriate tunneling properties in such devices. The efficiency of the whole tandem device under AM1.5 illumination is derived and optimized with respect to the halide composition and the thickness of the HOP absorbing top sub-cell. The optimum configuration ensuring current matching is obtained for a perovskite layer thickness ranging between 300 and 400 nm and a bromide concentration of 20%, which corresponds to a bandgap energy of 1.7 eV. The corresponding optimal tandem cell photovoltaic efficiency amounts to 27%, exceeding significantly the efficiencies of stand-alone silicon and perovskite cells taken for our simulations, which amount to 17.3% and 17.9%, respectively, and bypassing the record efficiency of crystalline Si cells. This study provides clear evidence that the perovskite methylammonium lead iodide-bromide $CH_3NH_3PbI_{3(1-x)}Br_{3x}$ alloy is a very promising absorbing material for high efficiency and low cost silicon based tandem solar cells.

**ACKNOWLEDGMENTS**

J. Even work is supported by the "Fondation d'entreprises banque Populaire de l'Ouest" under Grant PEROPHOT 2015. The work was supported by Cellule Energie du CNRS (SOLHYBTRANS Project) and University of Rennes 1 (Action Incitative, Défis Scientifiques Emergents 2015).

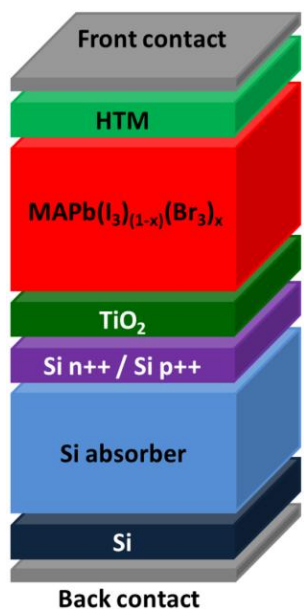

FIG. 1. (Color online) Overview of the structure showing the different layers taken into account in the simulation.

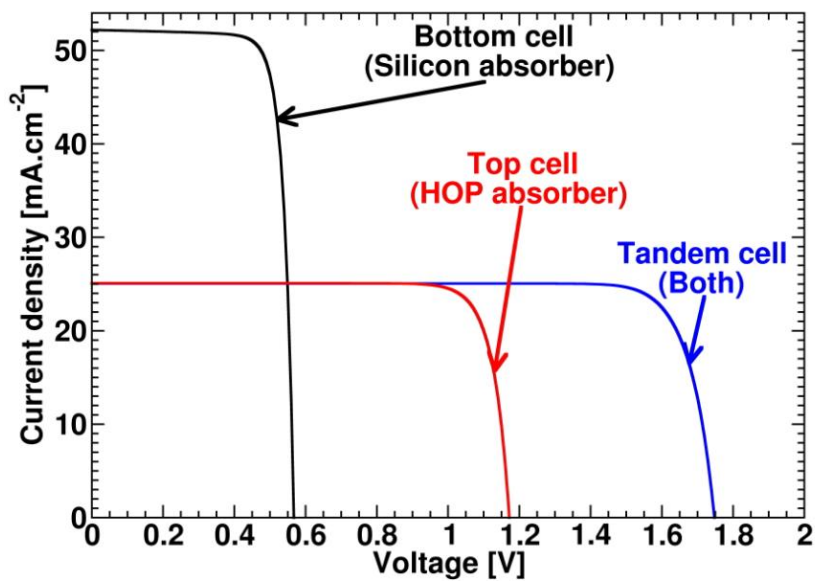

FIG. 2. (Color online) J-V characteristics of the bottom cell (silicon absorber, black line), top cell (HOP absorber, red line) and tandem cell (silicon and HOP absorbers, blue line) under AM1.5 illumination. The HOP absorber material is $CH_3NH_3PbI_{3(1-x)}Br_{3x}$ with x=0.2 and a thickness of 300 nm.

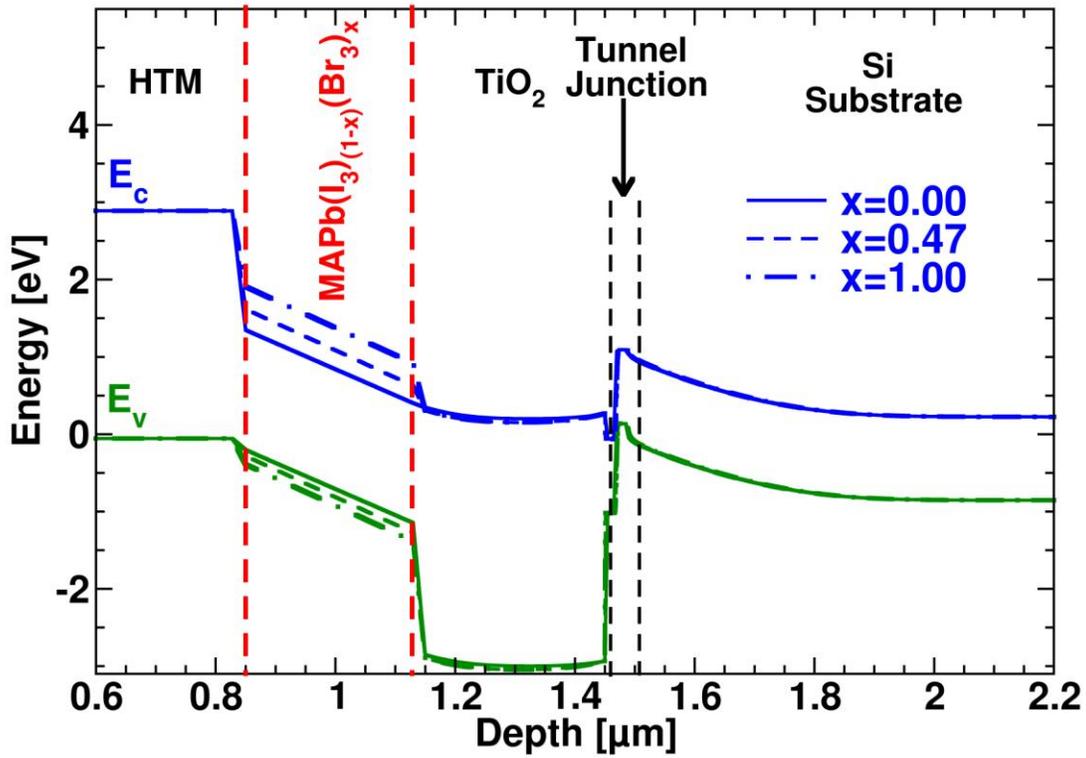

FIG. 3. (Color online) Energy Band diagrams of the tandem cell. The valence and conduction band edges (resp. $E_V$ and $E_C$) are depicted in green and blue, respectively. The position of the top cell is highlighted in red. Effect of halide composition in the HOP absorber $CH_3NH_3PbI_{3(1-x)}Br_{3x}$ is illustrated for x=0.0 (solid line), 0.47 (dashed line) and 1.0 (dashed dotted line). The thickness is kept constant at 300nm.

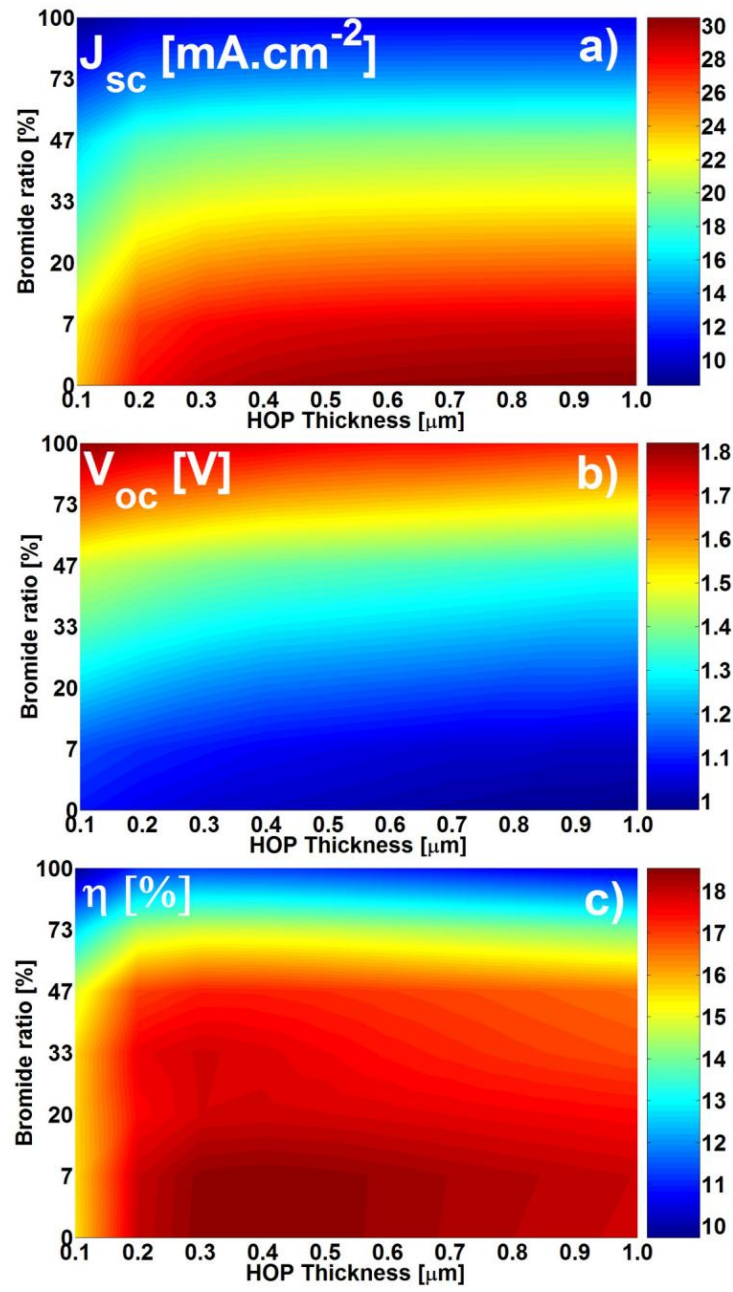

FIG. 4. (Color online) HOP top cell (a) short-circuit current $J_{sc}$, (b) open-circuit voltage $V_{oc}$, and (c) efficiency $\eta$.

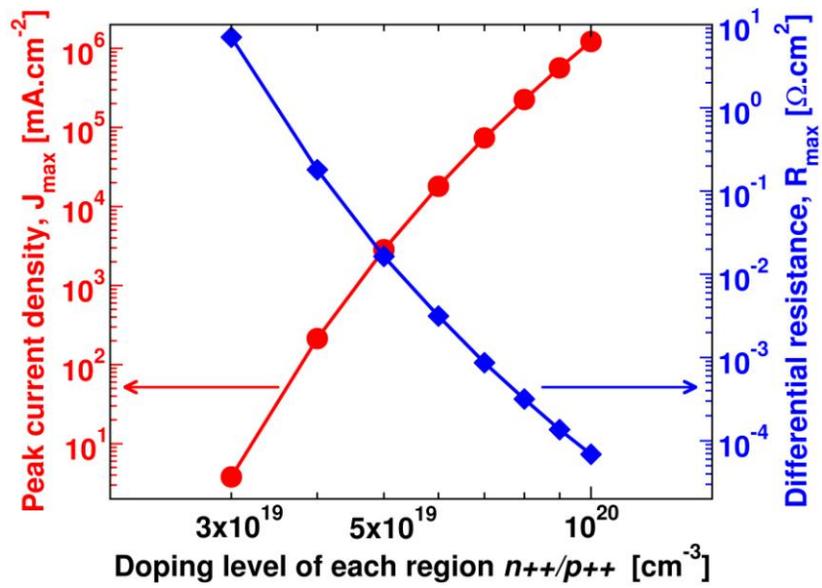

FIG. 5. (Color online) Characteristic of the tunnel junction (red dots) as a function of the doping. The peak current density $J_{max}$ [mA.cm$^{-2}$] and the differential resistance $R_{max}$ [$\Omega$.cm$^2$] are marked by red circles and blue squares, respectively.

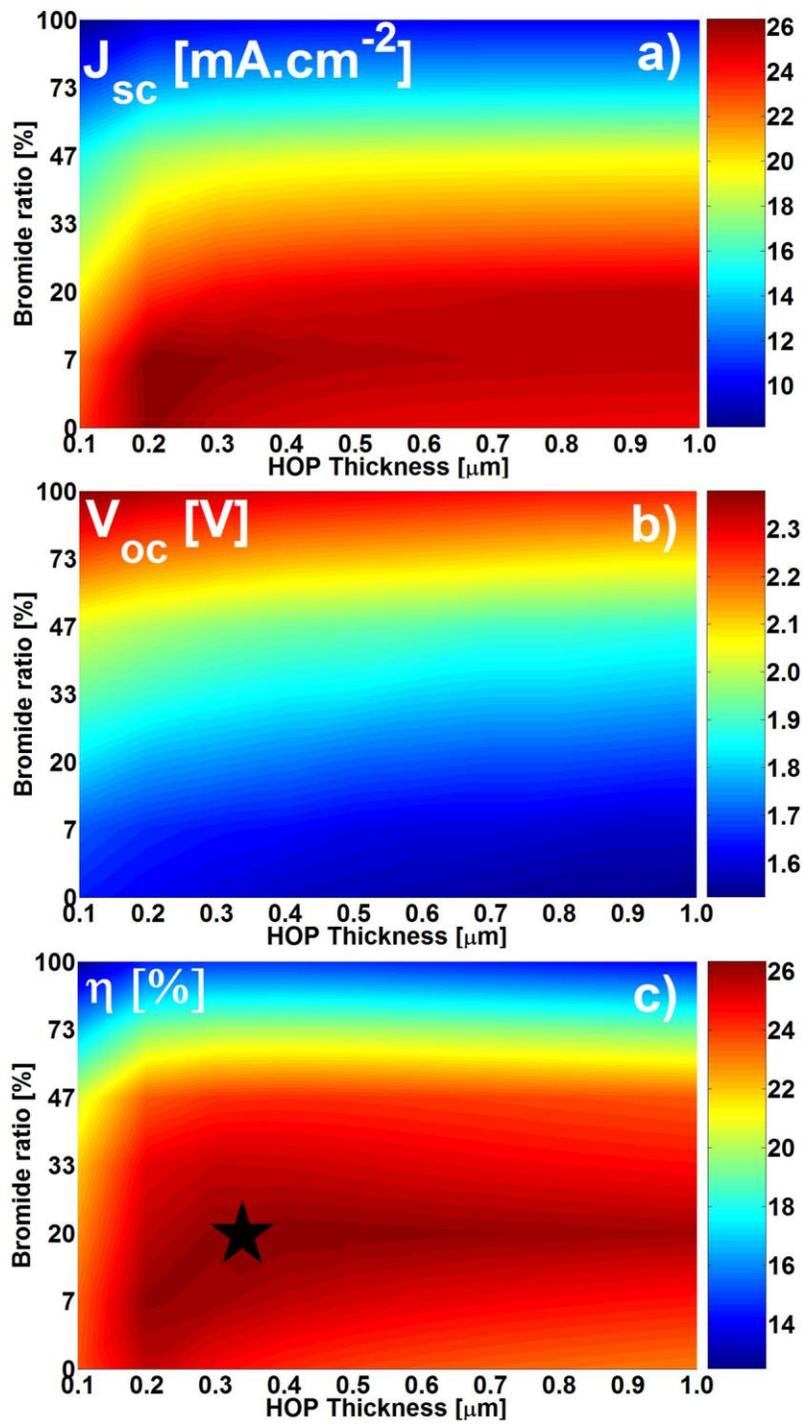

FIG. 6. (Color online) Tandem cell (a) short-circuit current $J_{sc}$, (b) open-circuit voltage $V_{oc}$, and (c) efficiency $\eta$. The black star in (c) marks the maximum efficiency 27% obtained for a bromide ratio of 20% and a thickness of 350 nm.

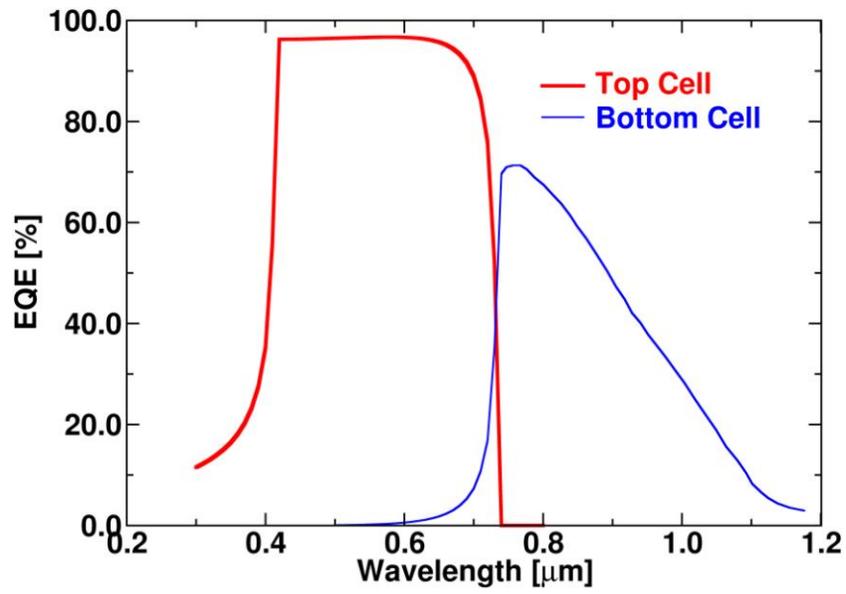

FIG. 7. (Color online) EQE of the top cell (red line) with alloy perovskite (x=0.2 and 300 nm of thickness), and EQE of the Si bottom cell (blue line).

TABLE I. Parameters used for the device simulations. $N_A$ and $N_D$ are the acceptors and donors doping concentrations, $N_C$ and $N_V$ are the effective densities of states in the conduction and valence band, $\varepsilon_r$ the relative permittivity, $\chi$ the electron affinity, $E_g$ the bandgap energy, $\mu_n$ and $\mu_p$ the electron and hole mobility, and $\tau_n$ and $\tau_p$ the electron and hole lifetime.

| Parameter | $CH_3NH_3PbI_{3(1-x)}Br_{3x}$ | HTM | $TiO_2$ | Si |
|---|---|---|---|---|
| $E_g$ (eV) | 1.55+0.75x | 3.0 | 3.2 | 1.12 |
| $\chi$ (eV) | 3.9-0.55x | 2.45 | 3.9 | 4.17 |
| $\varepsilon_r$ | 6.5 | 3.0 | 9 | 11.8 |
| $\mu_n/\mu_p$ (cm$^2$V$^{-1}$s$^{-1}$) | 2/2 | $2\,10^{-4}/2\,10^{-4}$ | 20/10 | 1500/480 |
| $\tau_n/\tau_p$ (s) | $10^{-6}/10^{-6}$ | $10^{-7}/10^{-7}$ | $10^{-7}/10^{-7}$ | $10^{-4}/10^{-4}$ |
| $N_C$ (cm$^{-3}$) | $2.2\,10^{18}$ | $2.2\,10^{18}$ | $2.2\,10^{18}$ | $2.8\,10^{19}$ |
| $N_V$ (cm$^{-3}$) | $1.8\,10^{19}$ | $1.8\,10^{19}$ | $1.8\,10^{19}$ | $1.04\,10^{19}$ |
| $N_A$ (cm$^{-3}$) | - | $2\,10^{18}$ | - | - |
| $N_D$ (cm$^{-3}$) | $10^{13}$ | - | $10^{16}$ | - |

TABLE II. Tandem cell performances for different compositions of the absorber $CH_3NH_3PbI_{3(1-x)}Br_{3x}$ for thickness of 300 nm and 400 nm. Traps effects are included. The corresponding values of band gap energy $E_g$ and electronic affinities $\chi$ are also reported (linear interpolation).

| | 300nm | | | | | | | 400nm | | | | | | |
|---|---|---|---|---|---|---|---|---|---|---|---|---|---|---|
| x | 0.00 | 0.07 | 0.20 | 0.33 | 0.47 | 0.73 | 1 | 0.00 | 0.07 | 0.20 | 0.33 | 0.47 | 0.73 | 1 |
| Eg (eV) | 1.55 | 1.6 | 1.7 | 1.8 | 1.9 | 2.1 | 2.3 | 1.55 | 1.6 | 1.7 | 1.8 | 1.9 | 2.1 | 2.3 |
| $\chi$ (eV) | 3.90 | 3.86 | 3.79 | 3.71 | 3.64 | 3.49 | 3.35 | 3.90 | 3.86 | 3.79 | 3.71 | 3.64 | 3.49 | 3.35 |
| $J_{sc}$ (mAcm$^{-2}$) | 25.3 | 26.2 | 24.4 | 21.4 | 18.6 | 13.8 | 9.9 | 24.9 | 25.8 | 24.8 | 21.7 | 18.9 | 14.0 | 10.0 |
| $V_{oc}$ (V) | 1.60 | 1.64 | 1.75 | 1.86 | 1.96 | 2.15 | 2.33 | 1.58 | 1.63 | 1.73 | 1.84 | 1.94 | 2.13 | 2.31 |
| FF (%) | 84.8 | 83.6 | 85.1 | 88.6 | 91.0 | 92.2 | 89.2 | 85.1 | 84.2 | 84.9 | 88.0 | 90.8 | 91.9 | 88.5 |
| $\eta$ (%) | 24.8 | 26.0 | 26.4 | 25.5 | 24.1 | 19.9 | 14.8 | 24.3 | 25.6 | 26.6 | 25.4 | 24.2 | 19.9 | 14.8 |